\documentclass[onecolumn]{aastex62}

\newcommand{\psr}{J2032}
\newcommand{\cxo}{\emph{Chandra}}
\newcommand{\xmm}{\emph{XMM-Newton}} 
\newcommand{\nus}{\emph{NuSTAR}} 
\newcommand{\sw}{\emph{Swift}} 
\newcommand{\ergs}{\,erg\,s$^{-1}$}
\newcommand{\ergcm}{\,erg\,cm$^{-2}$\,s$^{-1}$}

\newcommand{\ujb}{\,$\mu$Jy\,beam$^{-1}$}

\newcommand{\nh}{$N_{\rm H}$}
\newcommand{\cms}{\,cm$^{-2}$}
\newcommand{\kms}{\,km\,s$^{-1}$}

\begin{document}
\title{X-ray and Radio Variabilities of PSR J2032+4127 near Periastron}
\author[0000-0002-5847-2612]{C.-Y.~Ng}
\affil{Department of Physics, The University of Hong Kong, Pokfulam
Road, Hong Kong}
\email{ncy@astro.physics.hku.hk}

\author[0000-0002-6089-6836]{W. C. G. Ho}
\affil{Department of Physics and Astronomy, Haverford College, 370 Lancaster
Avenue, Haverford, PA 19041, USA}
\affil{Mathematical Sciences, Physics and Astronomy, and STAG Research
Centre, University of Southampton, Southampton SO17 1BJ, United Kingdom}

\author[0000-0003-3847-3957]{E. V. Gotthelf}
\affil{Columbia Astrophysics Laboratory, Columbia University, 550 West 120th
Street, New York, NY 10027-6601, USA}

\author[0000-0003-4814-2377]{J. P. Halpern}
\affil{Columbia Astrophysics Laboratory, Columbia University, 550 West 120th
Street, New York, NY 10027-6601, USA}

\author[0000-0002-0763-8547]{M. J. Coe}
\affil{Physics and Astronomy, University of Southampton, Southampton SO17
1BJ, UK}

\author[0000-0001-9242-7041]{B. W. Stappers}
\affil{Jodrell Bank Centre for Astrophysics, School of Physics and Astronomy, University of Manchester, Manchester M13 9PL, UK}

\author[0000-0002-4799-1281]{A. G. Lyne}
\affil{Jodrell Bank Centre for Astrophysics, School of Physics and Astronomy, University of Manchester, Manchester M13 9PL, UK}

\author{K. S. Wood}
\affil{Praxis Inc., Resident at the Naval Research Laboratory, Washington, DC 20375, USA}

\author[0000-0002-0893-4073]{M. Kerr}
\affil{Space Science Division, Naval Research Laboratory, Washington, DC 20375-5352, USA}

\correspondingauthor{C.-Y.~Ng}
\shorttitle{Variability of \psr\ near Periastron}
\shortauthors{Ng et al.}
\submitjournal{ApJ}

\keywords{stars: neutron
--- pulsars: individual (PSR J2032+4127)
--- stars: individual (MT91 213)
--- X-rays: binaries
--- X-rays: individual: PSR J2032+4127
}
\begin{abstract}
We present X-ray and radio monitoring observations of the gamma-ray binary PSR
J2032+4127/MT91 213 during its periastron passage in late 2017. Dedicated
\cxo, \xmm, \nus\ X-ray observations and VLA radio observations of this long
orbit (50\,years), 143\,ms pulsar/Be star system clearly revealed flux and
spectral variability during the passage. The X-ray spectrum hardened near
periastron, with a significant decrease in the power-law photon index from
$\Gamma\approx2$ to 1.2 and evidence of an increased absorption column
density. We identified a possible spectral break at a few keV in the spectrum
that suggests synchrotron cooling. A coincident radio and X-ray flare occurred
one week after periastron, which is possibly the result of the pulsar wind
interacting with the Be stellar disk and generating synchrotron radiation.
However, a multi-wavelength comparison indicate that the X-ray and radio
spectra cannot be simply connected by a single power-law component. Hence, the
emission in these two energy bands must originate from different particle
populations.
\end{abstract}

\section{Introduction}
Gamma-ray binaries are an emerging class of interacting binaries which emit
most of their radiation power in GeV and TeV energies \citep[see][for
reviews]{dub13,dgp+17}. These systems offer a unique perspective to study 
non-thermal emission processes at high energies. Most importantly,
their variabilities along the binary orbit give a powerful diagnostic tool for
the physical conditions of the outflows and the environment, providing
dynamical and geometrical information that is otherwise not possible from
isolated sources. 

There are fewer than ten gamma-ray binaries known and only two of them, PSRs
J2032+4127 and B1259$-$63, have pulsations detected, which firmly established
the pulsar nature of the compact objects. The pulsar spin parameters give
direct constraints on the energy input from its wind, allowing detailed
physical modeling \citep[e.g.,][]{ttn+17,pvc+18,cos+19}. \object{PSR~J2032+4127}
(hereafter \psr) was first identified as a gamma-ray pulsar with the
\emph{Fermi Gamma-Ray Space Telescope} \citep{aaa+09a} and then later detected
in radio and X-rays \citep{crr+09}. It has a spin period of 143\,ms and is
associated with the TeV source J2032+4130 \citep{aab+14}. Long-term radio
timing revealed a highly eccentric ($e=0.98$) orbit of the pulsar around a Be
star \object{MT91~213} in Cygnus OB2 association, with a binary period of
52.4\,yr (\citealp{lsk+15,hnl+17}; Stappers et al.\ in prep.). Gaia parallax
measurement of the companion star suggest a distance of
$1.39^{+0.08}_{-0.06}$\,kpc \citep{jkc+18}.

\psr\ was predicted to be at periastron on 2017 Nov 13 (MJD 58070.73)
(\citealp{hnl+17,csh+17}; Stappers et al.\ in prep.), and it has been closely
monitored at different wavelengths before and after that. In X-rays, we found
a flux increase by $\sim$20 times from 2002 to 2016 and a rapid brightening
since late 2015 \citep{hnl+17}. It was suggested that the brightening could be
attributed to strong X-ray flares on weekly to monthly timescales with a
slowly increasing baseline \citep{lkt+17}. During the periastron passage, the
X-ray flux showed a sharp dip which was then followed by a 20\,day long flare
\citep{ltn+18,cos+19}. Since the X-ray emission is believed to originate from
the interaction between the stellar wind and the pulsar wind, the flux
variation could indicate changes in the wind conditions, such as magnetization
\citep{ttn+17}, or could be due to geometrical effects \citep{pvc+18}. In
gamma-rays, \psr\ showed a different behavior which is difficult to explain.
The TeV flux increased by an order of magnitude and peaked at periastron, then
dropped abruptly one week later, and the low state lasted for a few days
\citep{abb+18}. The GeV flux, however, exhibited no variability over the same
period \citep{ltn+18}. All these are not well modeled by current theories
\citep[e.g.,][]{ttn+17,bbs18}.

In this paper, we report on our monitoring campaign of \psr\ in X-rays and
radio, using the \emph{Chandra X-ray Observatory}, \xmm, \nus, and the Jansky
Very Large Array (VLA). Compared to the \sw\ X-ray results reported in
previous studies, the sensitive instruments we used offered much better
constraints on the spectral parameters, enabling us to determine not only the
light curve but also spectral variabilities of the source near periastron.

\section{Observations and Data Analysis}
\subsection{\emph{Chandra X-ray Observatory}}
We carried out new \cxo\ ACIS-S observations between 2017 Feb to 2018 Oct in
timed exposure mode with 1/4 subarray. This gave a frame time of 0.8\,s such
that pileup is not an issue ($\lesssim3\%$ even when the source was
brightest). The observation parameters are listed in Table~\ref{tab:xobs}.
Data reduction was performed with CIAO 4.10 and CALDB 4.8.1. We first
reprocessed the data using the task \texttt{chandra\_repro}, then checked that
there was no background flaring during the exposure. We generated light curves
of \psr\ but did not find any significant short-term variability between the
frame time and the exposure time, i.e.\ for a timescale of $\sim$1\,s to a few
$10^4$\,s.

\begin{deluxetable}{cccDDDDDD}
\tablewidth{0pt}
\tablecaption{X-ray Observations of \psr\ Used in this Study \label{tab:xobs}}
\tablehead{\colhead{Obs.\ Date}& \colhead{ObsID}& \colhead{MJD} &
\multicolumn2c{True Anomaly} & \multicolumn2c{Net Exposure} &
\multicolumn2c{Net Count} &
\multicolumn2c{$f_{p, N=1}^{u.l.}$\tablenotemark{b}} &
\multicolumn2c{$f_{p, N=1000}^{u.l.}$\tablenotemark{b}} \\
&&&\multicolumn2c{(\arcdeg)}& \multicolumn2c{(ks)}& \multicolumn2c{Rate\tablenotemark{a}
(ks$^{-1}$)} & \multicolumn2c{(\%)}& \multicolumn2c{(\%)}}
\decimals
\startdata
\sidehead{\cxo}
2017 Feb\phantom{;} 01 & 19607 &  57785.4&27.2 & 26.8 & 66.8 & \nodata & \nodata \\
2017 May 28 & 19700 & 57901.8 &33.6  & 28.6 & 143.9 & \nodata & \nodata \\
2017 Aug 29 & 19701 & 57994.2 &46.0  & 28.6 & 129.0 & \nodata & \nodata \\
2017 Nov 03 & 19702 & 58060.4 &102.1 & 19.1 & 114.8 & \nodata & \nodata \\
2017 Nov 05 & 20836 & 58062.9 &110.7 & 9.6 & 98.9 & \nodata & \nodata \\
2018 Jan 09 & 19608 & 58128.2 &307.7 & 29.0 & 108.5 & \nodata & \nodata \\
2018 Mar 16 & 19698 & 58193.8 &321.8 & 29.2 & 35.4 & \nodata & \nodata \\
2018 Aug 19 & 20599 & 58349.6 &332.6 & 27.7 & 15.5 & \nodata & \nodata \\
2018 Oct 02 & 20848 & 58393.6 &334.2 & 26.3 & 4.1 & \nodata & \nodata \\
\sidehead{\xmm}                            
2017 Oct 23 & 0801910201 & 58050.1 & 79.9  & 22.3 & 488.2 & 4.9 & 7.2  \\
2017 Nov 05 & 0801910301 & 58063.0 & 111.0 & 16.4 & 261.2 & 8.0 & 11.7 \\
2017 Nov 13 & 0801910401 & 58070.7 & 178.0 & 10.9 & 241.2 & 10.2 & 15.0 \\
2017 Nov 22 & 0801910501 & 58079.3 & 252.1 & 13.9 & 551.4 & 5.8 & 8.5  \\
2017 Dec 04 & 0801910601 & 58091.9 & 280.9 & 10.9 & 413.5 & 7.6 & 11.2 \\
\sidehead{\nus}                            
2017 Oct 23 & 30302002002 & 58050.2 &80.1  & 37.9 & 182.1 & 6.5 & 9.5  \\
2017 Nov 05 & 90302321002 & 58062.9 &110.5 & 39.8 & 98.3  & 9.0 & 13.3 \\
2017 Nov 13 & 30302002004 & 58070.5 &177.7 & 42.7 & 82.7  & 9.6 & 14.2 \\
2017 Nov 21 & 90302321004 & 58079.2 &251.6 & 40.0 & 184.0 & 6.2 & 9.1  \\
2017 Dec 04 & 30302002006 & 58092.2 &281.2 & 41.5 & 130.7 & 7.6 & 11.1
\enddata                                   
\tablenotetext{a}{In energy range of 0.5--8\,keV for \cxo, 0.5--10\,keV for
\xmm, and 2--79\,keV for \nus. The \xmm\ values are for the PN only,
and the \nus\ values are the sum from the FPMA and FPMB detectors.}
\tablenotetext{b}{$3\sigma$ upper limits on the pulsed fraction, assuming a
sinusoidal signal. $f_{p, N=1}^{u.l.}$ and $f_{p, N=1000}^{u.l.}$ correspond
to 1 and 1000 search trials, respectively. }
\end{deluxetable}

We extracted the spectrum of \psr\ from a 5\arcsec\ radius aperture and the
background spectrum from nearby source-free regions. A large aperture was
chosen to enclose most of the source flux, since the background is generally
negligible. We grouped the spectra to at least 20 counts per bin, except for
the last observation, which was grouped at 15 due to low count rate, and
performed spectral analysis using the \emph{Sherpa} package. We fit the
pulsar spectra in the 0.5--8\,keV range with an absorbed power-law (PL) model.
The absorption cross section and element abundances given by \citet{wam00} are
adopted. Finally, uncertainties in flux are estimated with Monte Carlo
simulations.

\subsection{\xmm\ and \nus}
In addition to \cxo, we obtained five coordinated \xmm\ and \nus\ observations
of \psr, taken on the day of periastron passage, and one and three weeks
before and after that. The observation parameters are listed in
Table~\ref{tab:xobs}. For \xmm, the PN camera operated in the small window
mode with a high time resolution of 6\,ms, while the MOS cameras were in the
large window mode with 0.9\,s time resolution. The data reduction was
performed with SAS v.15 and the most up-to-date calibration files. We followed
the standard pipeline processing procedure and then filtered out time periods
with high background. The resulting exposure times are list in
Table~\ref{tab:xobs}. The \nus\ observations have a time resolution of better
than 3\,ms and the data reduction was performed using the standard processing
pipeline in the latest version of NuSTARDAS.

For timing analysis, we extracted \xmm\ counts in the 0.5--10\,keV range from
a 0\farcm5 radius aperture and \nus\ counts in 20--30\,keV from a
1\arcmin\ radius aperture to generate light curves. The source count rates
are found to be nearly constant over an observation and shorter time
scales, with rms fluctuations of less than two times the mean statistical error
(i.e.\ rms $\leq2\bar\sigma$). This is confirmed by FFTs of each data set that
show little evidence of red noise or enhanced white noise.
We also performed a pulsation search using \xmm\ PN and \nus\ events. Photon
arrival times were converted to barycentric dynamical time (TDB) using the
DE405 solar system ephemeris. The arrival times were further adjusted to
account for binary motion of the pulsar. We searched for periodic signals over
a small range of frequency near the pulsar period from radio timing
(Stappers et al.\ in prep.). We tried the $Z_1^2$ test and also the $\chi^2$
test with 10 phase bins, but no signals were found. The $3\sigma$ upper limits
on the pulsed fraction for a putative sinusoidal signal are given in
Table~\ref{tab:xobs}, after taking into account the background estimate and
the number of search trials.

We extracted the \xmm\ spectra of \psr\ from a circular aperture of 0\farcm5
radius. There are four background X-ray sources in the aperture (see
Figure~\ref{fig:cxo}), but the contamination should be negligible, given
their low count rates (see Appendix~\ref{sec:bg}). The MOS background was
extracted from an annular region with radii between 1\arcmin\ and 2\arcmin\
centered on the pulsar, and the PN background was from a 1\arcmin\ radius
circle offset by 1\farcm7 along the CCD readout direction. The \nus\ source
and background spectra were extracted from a 0\farcm8 radius circular region
and an annular region between 1\farcm2 and 1\farcm9, respectively. We merged
the \xmm\ MOS spectra taken in the same day for a joint analysis, and did the
same for the two \nus\ FPM detectors. The source spectra were grouped with a
minimum of 100 counts per bin and then fit using the \emph{XSPEC} package.

\begin{figure}[ht]
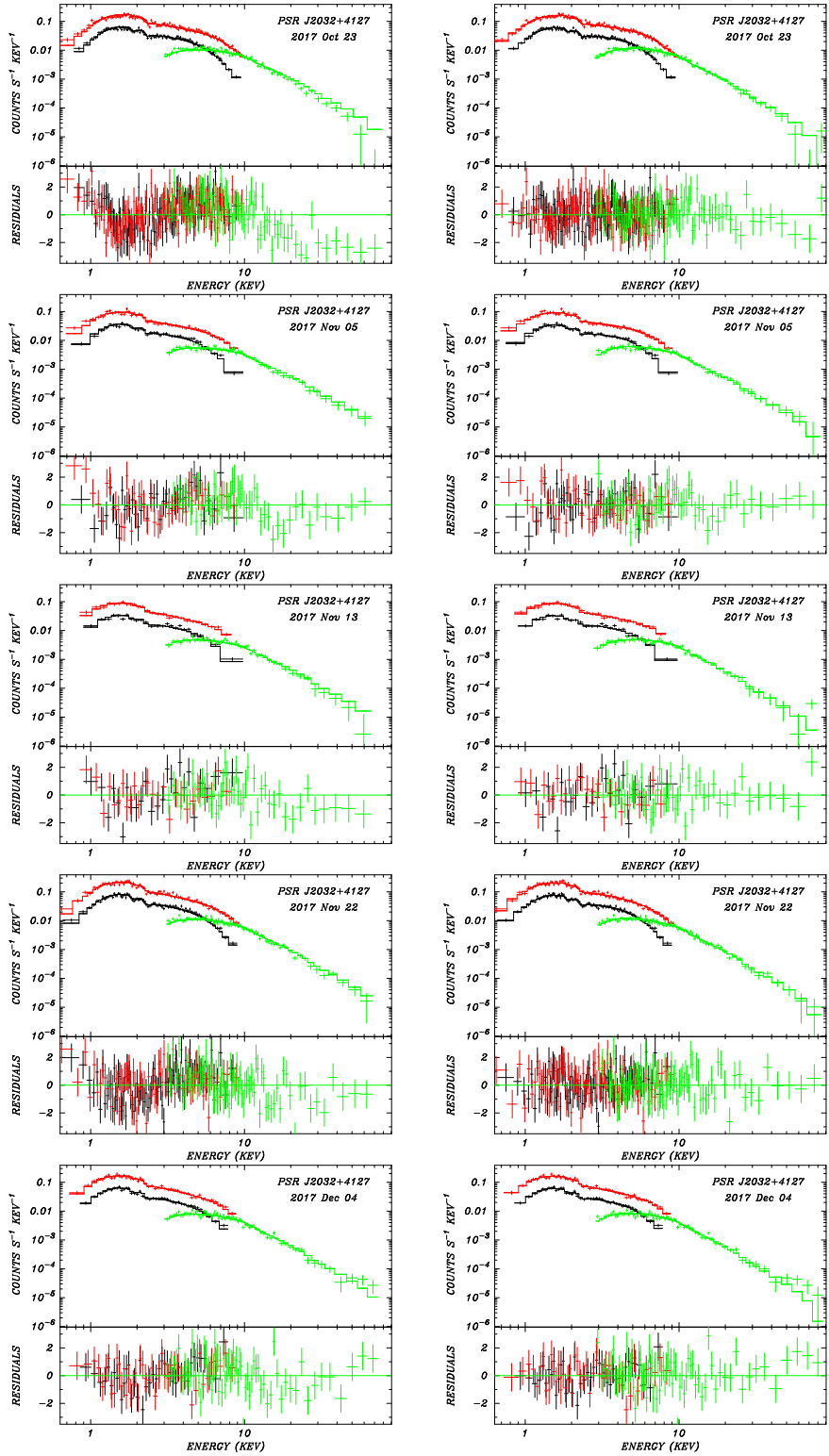

\centering
\epsscale{0.6}
\includegraphics[height=0.3\textwidth,angle=270]{j2032_spec_pow_grp1.eps} \hspace*{5mm}
\includegraphics[height=0.3\textwidth,angle=270]{j2032_spec_bknpow_grp1.eps}\\
\includegraphics[height=0.3\textwidth,angle=270]{j2032_spec_pow_grp2.eps} \hspace*{5mm}
\includegraphics[height=0.3\textwidth,angle=270]{j2032_spec_bknpow_grp2.eps}\\
\includegraphics[height=0.3\textwidth,angle=270]{j2032_spec_pow_grp3.eps} \hspace*{5mm}
\includegraphics[height=0.3\textwidth,angle=270]{j2032_spec_bknpow_grp3.eps}\\
\includegraphics[height=0.3\textwidth,angle=270]{j2032_spec_pow_grp4.eps} \hspace*{5mm}
\includegraphics[height=0.3\textwidth,angle=270]{j2032_spec_bknpow_grp4.eps}\\
\includegraphics[height=0.3\textwidth,angle=270]{j2032_spec_pow_grp5.eps} \hspace*{5mm}
\includegraphics[height=0.3\textwidth,angle=270]{j2032_spec_bknpow_grp5.eps}
\caption{Joint fits to the \xmm\ and \nus\ spectra of \psr, with an absorbed
single PL model (left) and an absorbed broken PL model (right). \label{fig:xmmnus}}
\end{figure}
We first fit the \xmm\ spectra using the same model and energy range as the
\cxo\ analysis above for a direct comparison. We then performed a joint fit
to the \xmm\ and \nus\ spectra in the 0.5--10\,keV and 3--79\,keV energy
ranges, respectively. An absorbed PL model was employed. We linked the column
density (\nh) and the PL photon index ($\Gamma$) of all spectra at each epoch,
but allow different PL normalizations to account for any cross-instrument flux
mis-calibrations. The fitting results are formally acceptable with reduced
$\chi^2$ values from 1.1 to 1.5. The normalizations between different
instruments are well consistent to within 5\%. The best-fit spectra and
residuals are shown in the left panel of Figure~\ref{fig:xmmnus}. There is a
hint of systematic trend in the residuals, suggesting that a more complex
spectral model may be needed. We therefore tried a broken PL model. This
improves the fits and gives lower reduced $\chi^2$ values of 0.9--1.1. The
results are plotted in the right panels of Figure~\ref{fig:xmmnus}. The broken
PL fit clearly reduces the residuals, particularly at low energy. We performed
an $F$-test and confirm that the additional component is statistically
significant.

\subsection{VLA}
We carried out six new radio observations of \psr\ with the VLA. Except
the first one, the remaining five were taken within a couple days of the \xmm\ and
\nus\ observations. The VLA was in the B array configuration, and the data
were taken in the L, S, C, and X bands, covering a frequency from 1 to
10\,GHz. In addition, we analyzed archival data observed earlier
in 2017 in the S band with the C array configuration. The observation
parameters are listed in Table~\ref{tab:robs}. The time resolution was 3\,s in
all data, too low to resolve the pulsar spin period. Standard
calibrators, either 3C48, 3C138, or 3C286 were observed for absolute flux
density scale, delay, and bandpass calibrations. A nearby phase calibrator
J2007+4029 was observed for gain calibration. We also observed J1407+2827 and
3C147 for polarization calibration. Note that for the 2017 Nov
13 observation, 3C147 was used as the primary calibrator since the 3C138 data were
severely contaminated by RFI.

\begin{deluxetable}{cccccccc}
\tablewidth{0pt}
\tablecaption{VLA Observations of \psr\ Used in this Study \label{tab:robs}}
\tablehead{\colhead{Obs.\ Date}&\colhead{MJD} &
\colhead{Array}& \colhead{Band} & \colhead{No.\ of} & \colhead{Center} &
\colhead{Bandwidth} & \colhead{Integration} \\
& & & & \colhead{Antennas} & \colhead{Freq.\ (GHz)} &
\colhead{(GHz)} & \colhead{Time (min)}} 
\startdata
2017 Aug 14 & 57979.2 & C & S & 27 & 3 & 2 & 23.7 \\
2017 Sep 03 & 57999.4 & B & L, S, C, X & 27 & 1.52, 3, 5.5, 9 & 1, 2, 2, 2 & 7.0, 3.7, 3.7, 3.7 \\
2017 Oct 24 & 58050.1 & B & L, S, C, X & 25 & 1.52, 3, 5.5, 9 & 1, 2, 2, 2 & 7.0, 3.8, 3.6, 3.6 \\
2017 Nov 07 & 58064.1 & B & L, S, C, X & 27 & 1.52, 3, 5.5, 9 & 1, 2, 2, 2 & 6.9, 3.8, 3.6, 3.6 \\
2017 Nov 13 & 58070.1 & B & L, S, C, X & 25 & 1.52, 3, 5.5, 9 & 1, 2, 2, 2 & 7.0, 3.7, 3.7, 3.7 \\
2017 Nov 21 & 58078.9 & B & L, S, C, X & 27 & 1.52, 3, 5.5, 9 & 1, 2, 2, 2 & 6.9, 4.9, 4.2, 4.2 \\
2017 Dec 03 & 58090.1 & B & L, S, C, X & 26 & 1.52, 3, 5.5, 9 & 1, 2, 2, 2 & 7.0, 3.7, 3.7, 3.7\enddata
\end{deluxetable}

All data reduction was done with CASA 5.1. We first applied the VLA
calibration pipeline for basic flagging and calibration, then performed
additional flagging when needed. We formed the radio images and carried out
deconvolution with the task \texttt{tclean}. Briggs weighting with parameter
\emph{robust}=0.5 was used to form L and S band images, while natural
weighting was used for C and X bands. These weights were chosen to optimize
the signal-to-noise ratio to boost the detection sensitivity. The final
intensity maps have rms noise levels of $\sim$50\ujb\ in the L band,
$\sim$20\ujb\ in the S band, and $\sim$15\ujb\ in C and X bands. The beam
size ranges from $\sim4''$ FWHM in the L band to $\sim1''$ FWHM in the X band.
Finally, we generated polarization images for the 2017 Nov 21 observation,
when the pulsar was brightest. However, no polarization signal was found.

\section{Results}
%\subsection{Imaging and Source Variability}
\subsection{X-rays}
\begin{figure}[ht]
\centering
\epsscale{0.6}
\plotone{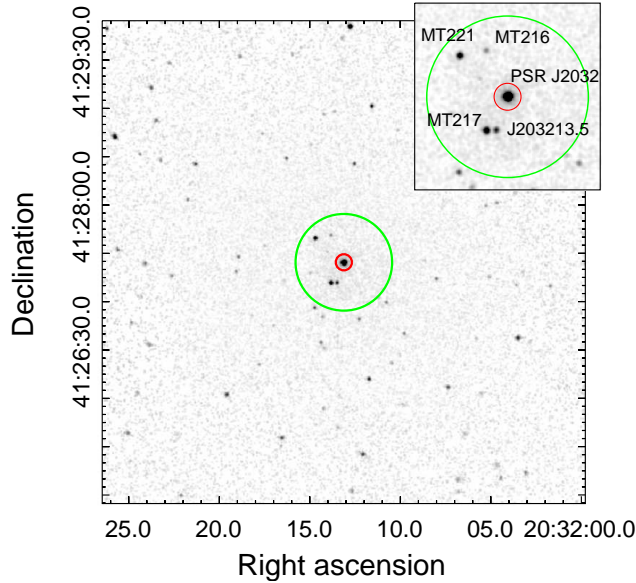}
\caption{Exposure-corrected \cxo\ X-ray image of the field centered on \psr\
in 0.5--7\,keV range, lightly smoothed to 2\arcsec\ resolution. The red and
green circles have radii of 5\arcsec\ and 30\arcsec\ radius that indicate the
extraction regions for \cxo\ and \xmm\ spectral analysis, respectively.
\emph{Inset:} zoom-in of the image showing four nearby background sources:
MT91 221, MT91 216, MT91 217, and CXOU J203213.5+412711 (see
Appendix~\ref{sec:bg}).
\label{fig:cxo}}
\end{figure}

Figure~\ref{fig:cxo} shows an exposure-corrected X-ray image of the field of
\psr, generated by co-adding all on-axis \cxo\ observations, including those
used in this study and in our previous work \citep{hnl+17}. There is no
obvious extended emission surrounding the pulsar. We plotted the radial count
profile of the pulsar and found that it is consistent with a model PSF
simulated by
ChaRT\footnote{\url{http://cxc.harvard.edu/ciao/PSFs/chart2/index.html}}.
We were unable to confirm the large scale ($\sim$1\farcm6 diameter) diffuse
emission reported in previous studies \citep{mgh07,crr+09}. However, we note
that many of the \cxo\ exposures were taken in the subarray mode with a small
field of view ($\sim$2\arcmin). \psr\ was also very bright during most of the
\cxo\ and \xmm\ observing epochs such that dust scattering could be a
potential issue. All these preclude a sensitive search for large scale
X-ray structure.

\psr\ was clearly detected in all X-ray observations. It exhibited significant
brightness variation between epochs. The best-fit spectral parameters to the
\cxo\ and \xmm\ spectra are listed in Table~\ref{tab:spec_cxoxmm}. The flux
obtained from different \xmm\ cameras are well matched to within 5\%. We
therefore report only the average values in the table. A simple absorbed PL
provides adequate fits to the X-ray spectra at different epochs, although the
parameters are not very well constrained in the last \cxo\ observation due to
the small number of counts. The \cxo\ and \xmm\ results are generally
consistent. In Table~\ref{tab:spec_xmmnus} we report results from the joint
\xmm\ and \nus\ spectral fits. As mentioned, the fits with a simple PL are
acceptable, but a broken PL gives better results and the improvement is
statistically significant as shown by an $F$-test. In the latter model, a
spectral break around 4 to 8\,keV was found and the PL photon index changes by
$\Delta\Gamma=0.3$--0.5 across the break. 

\begin{deluxetable}{lLLLLLL}
\tablewidth{0pt}
%\tabletypesize{\small}
\tablecaption{Single PL Fits to the \cxo\ and \xmm\ Spectrum of \psr
\label{tab:spec_cxoxmm}}
\tablehead{\colhead{Obs.\ Date}&\colhead{\nh}& \colhead{$\Gamma$} &
\colhead{Abs.\ Flux} & \colhead{$\chi^2_\nu$/dof} \\
& \colhead{($10^{22}$\cms)}& & \colhead{($10^{-13}$\ergcm)}}
\decimals
\startdata
\\[-10pt]
\multicolumn{5}{l}{\cxo\ (single PL)}\\[5pt]\hline
%\cutinhead{\cxo\ (single PL)}
2017 Feb 01 & 1.6\pm0.2 & 2.1\pm 0.2 & 9.6\pm0.6 & 1.12 / 73 \\
2017 May 28 & 1.5\pm0.1 & 1.7\pm0.1 & 24.7\pm1.1 & 0.78 / 157 \\
2017 Aug 29 & 1.3_{-0.1}^{+0.2} & 1.4\pm0.1 & 25.2^{+1.1}_{-1.2} & 0.91 / 145 \\
2017 Nov 03 & 1.4\pm0.2 & 1.2\pm0.1 & 25.5^{+1.5}_{-1.7} & 1.02 / 92 \\
2017 Nov 05 & 1.3\pm0.4 & 1.2\pm0.2 & 21.8^{+2.0}_{-2.2} & 1.17 / 42 \\
2018 Jan 09 & 1.5\pm0.2 & 1.6\pm0.1 & 20.0^{+1.0}_{-1.1} & 0.78 / 126 \\
2018 Mar 16 & 1.0_{-0.2}^{+0.3} & 1.6\pm0.2 & 6.1\pm0.6 & 0.83 / 45 \\
2018 Aug 19 & 1.1_{-0.4}^{+0.5} & 2.1_{-0.3}^{+0.4} & 2.0\pm0.3 & 1.29 / 18 \\
2018 Oct 02 & 1.0_{-0.8}^{+1.3} & 1.5_{-0.7}^{+0.9} & 0.6^{+0.2}_{-0.3} & 0.95 / 5 \\[5pt]\hline 
\\[-10pt]\multicolumn{5}{l}{\xmm\ (single PL)}\\[5pt]\hline
2017 Oct 23 & 1.02\pm0.05 & 1.19\pm0.04 & 31.0\pm0.6 & 1.0 / 186\\
2017 Nov 05 & 0.96_{-0.08}^{+0.09} & 1.24\pm0.06 & 16.5_{-0.6}^{+0.5} & 1.1 / 78 \\
2017 Nov 13 & 0.92_{-0.12}^{+0.13} & 1.29\pm0.09 & 14.6\pm0.6 & 1.1 / 45 \\
2017 Nov 22 & 1.02\pm0.05 & 1.41\pm0.04 & 32.4\pm0.7 & 1.2 / 149\\
2017 Dec 04 & 0.98_{-0.07}^{+0.08} & 1.43\pm0.06 & 24.3_{-0.8}^{+0.7} & 0.8 / 81
\enddata 
\tablecomments{All uncertainties are at 90\% confidence level. The best-fit
absorbed flux is in the 0.5--8\,keV energy range, and the \xmm\ flux are
the average values from all three cameras.}
\end{deluxetable}

\begin{deluxetable}{lLLLLLL}
\tablewidth{0pt}
%\tabletypesize{\small}
\tablecaption{Joint \xmm\ and \nus\ Fits to the Spectrum of \psr
\label{tab:spec_xmmnus}}
\tablehead{\colhead{Obs.\ Date}&\colhead{\nh}&
\colhead{$\Gamma/\Gamma_1$}&\colhead{$\Gamma_2$} &
\colhead{$E_{\rm break}$}& \colhead{Abs.\ Flux}&
\colhead{$\chi^2_\nu$/dof} \\
& \colhead{($10^{22}$\cms)}& & &
\colhead{(keV)} & \colhead{($10^{-13}$\ergcm)}}
\decimals
\startdata
\\[-10pt]\multicolumn{7}{l}{\xmm\ and \nus\ (single PL)}\\[5pt]\hline
%\sidehead{\xmm\ and \emph{NuSTAR} (single PL)}
2017 Oct 23 & 1.25\pm0.07 & 1.39\pm0.04&\nodata&\nodata& 34.0\pm0.4 & 1.45/273 \\
2017 Nov 05 & 1.13\pm0.10 & 1.40\pm0.06&\nodata&\nodata& 17.9\pm0.3 & 1.11/133 \\
2017 Nov 13 & 1.12\pm0.13 & 1.45\pm0.07&\nodata&\nodata& 15.6\pm0.3 & 1.16/95 \\
2017 Nov 22 & 1.16\pm0.06 & 1.54\pm0.04&\nodata&\nodata& 34.5\pm0.4 & 1.34/233 \\
2017 Dec 04 & 1.11\pm0.08 & 1.57\pm0.05&\nodata&\nodata& 25.3\pm0.4 & 1.18/151
\\[5pt]\hline
%\sidehead{\xmm\ and \emph{NuSTAR} (broken PL)}
\\[-10pt]\multicolumn{7}{l}{\xmm\ and \nus\ (broken PL)}\\[5pt]\hline
2017 Oct 23 &0.97\pm0.06&1.13\pm0.06&1.63\pm0.06&5.3\pm0.7&35.3\pm0.4&0.91/271 \\
2017 Nov 05 &0.91\pm0.09&1.18\pm0.07&1.55\pm0.09&4.7\pm0.8&18.4\pm0.3&0.92/131 \\
2017 Nov 13 &0.93\pm0.08&1.30\pm0.17&1.70\pm0.12&8.1\pm2.4&16.2\pm0.3&0.99/93 \\
2017 Nov 22 &0.93\pm0.11&1.26\pm0.06&1.65\pm0.08&3.8\pm1.0&35.2\pm0.4&1.10/231 \\
2017 Dec 04 &1.00\pm0.06&1.46\pm0.10&1.73\pm0.08&6.4\pm1.7&25.9\pm0.4&1.10/149
\enddata 
\tablecomments{All uncertainties are at 90\% confidence level. The best-fit
absorbed flux is the average values from different cameras in the 2--10\,keV
energy range.}
\end{deluxetable}

\begin{figure}[ht]
\centering
\epsscale{0.6}
\plotone{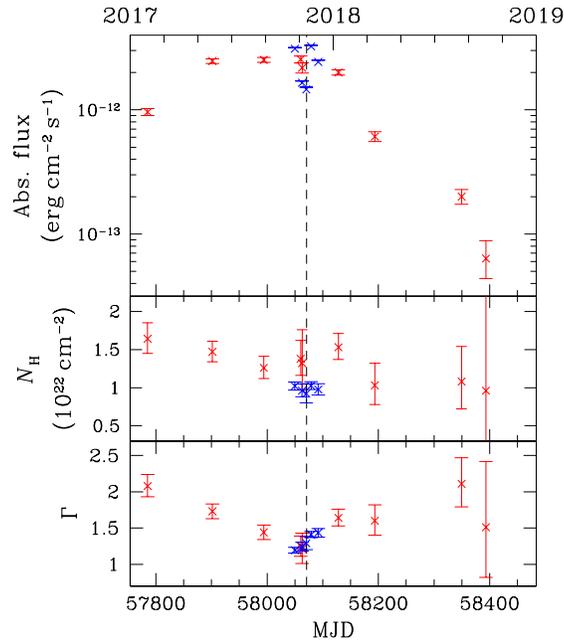}
\caption{Best-fit X-ray flux, column density (\nh), and photon index ($\Gamma$)
of \psr, obtained from the single PL model as listed in
Table~\ref{tab:spec_cxoxmm}. The red and blue data points indicate the \cxo\
and \xmm\ results, respectively. The top panel shows the absorbed flux in the
0.5--8\,keV energy range. The vertical dashed line marks the date of periastron
passage.
\label{fig:xspec_t}}
\end{figure}
\begin{figure}[ht]
\centering
\epsscale{0.6}
\plotone{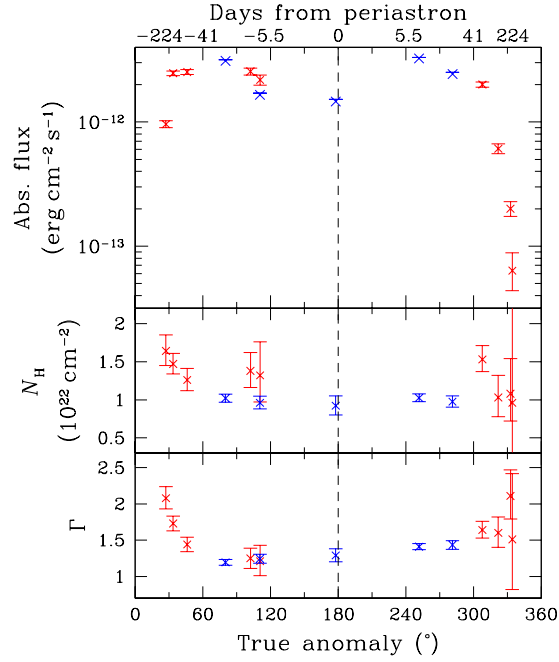}
\caption{Same as Figure~\ref{fig:xspec_t} but plotted against true anomaly.
\label{fig:xspec_phi}}
\end{figure}
\begin{figure}[ht]
\centering
\epsscale{0.6}
\plotone{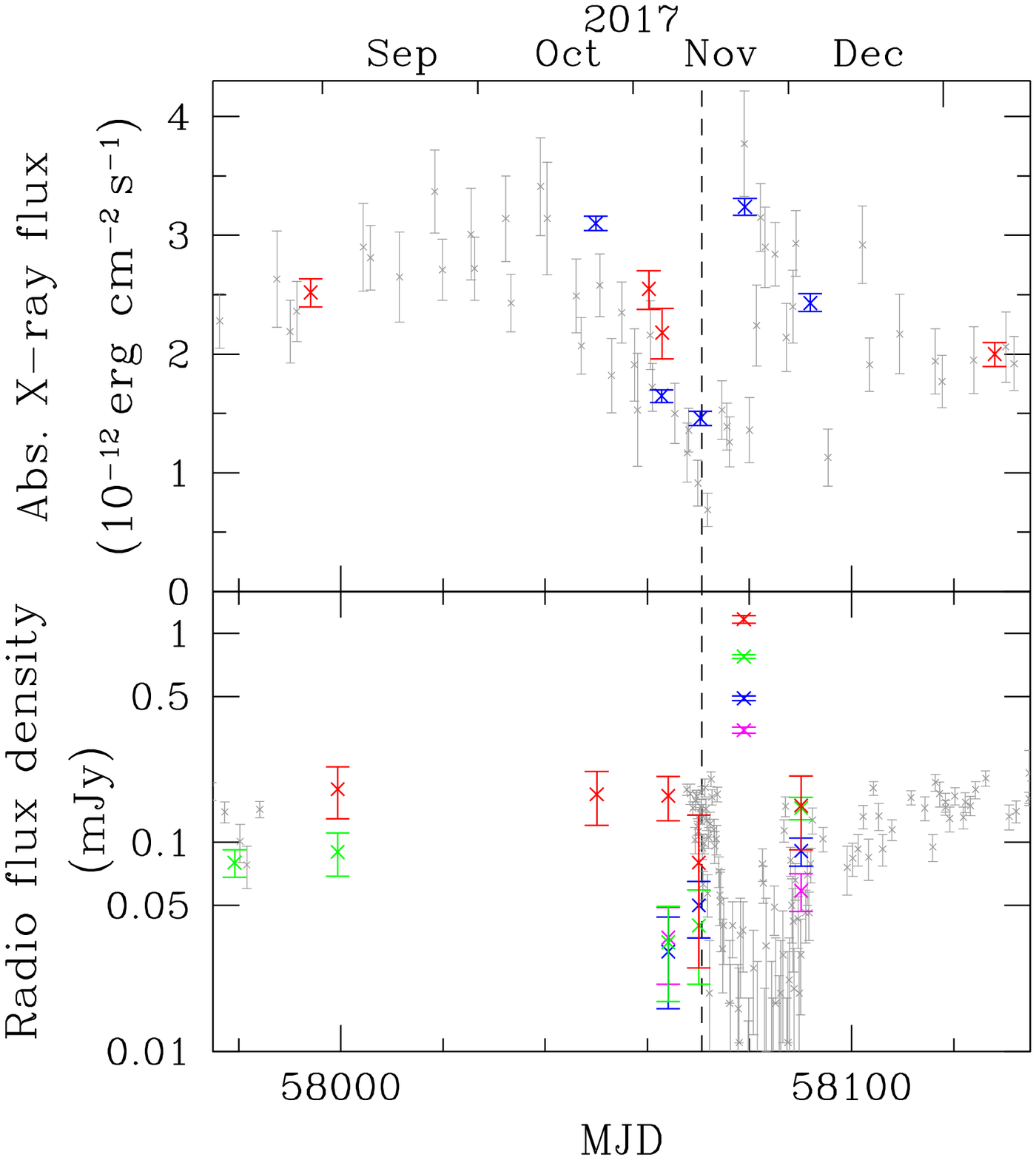}
\caption{\emph{Top:} X-ray flux of \psr\ as in Fig.~\ref{fig:xspec_t} (red: \cxo;
blue: \xmm+\nus) compared with the \sw\ light curve (grey) around periastron.
All these are absorbed flux in 0.5--8\,keV range. \emph{Bottom:} radio flux
densities of \psr\ over the same period. The VLA detections at the L, S, C,
and X bands are shown in red, green, blue, and magenta, respectively. The grey
data points indicate the pulsed flux density at the L band for comparison
(Stappers et al.\ in prep.). The vertical dashed line marks the date of
periastron passage. \label{fig:comp}}
\end{figure}

As we did not try fitting a broken PL to the \cxo\ data due to the low
sensitivity of the telescope at high energies, for the rest of the section, we
focus on the single PL fit in order to investigate the spectral evolution of
\psr. The best-fit spectral parameters at different epochs are plotted in
Figure~\ref{fig:xspec_t}. The column density shows a hint of a decreasing
trend prior to periastron, but after that, the variability is less clear due
to limited sensitivity of the measurements. On the other hand, the PL photon
index exhibited a more significant change. It decreased from $\Gamma\approx2$
in early 2017 to $\approx1.2$ around periastron, then rose back with a similar
timescale. The flux evolution is plotted in Figure~\ref{fig:xspec_t}. It
peaked at $3.5\times10^{-12}$\ergcm\ one week after periastron. After
correcting for interstellar absorption and scaling to a distance of 1.39\,kpc,
this value corresponds to a luminosity of $8.7\times10^{32}$\ergs\ in
2--10\,keV range. In Figure~\ref{fig:xspec_phi}, the spectral parameters are
plotted against the true anomaly, and they show similar evolution trends.

We can compare our results with the X-ray light curve from \sw. The latter was
converted from the count rates reported by \citet{cos+19}, assuming average
$N_{\rm H}=1.15\times10^{22}$\cms\ and $\Gamma=1.47$ from the \xmm\ and \nus\
fits. The results are plotted in Figure~\ref{fig:comp}. The flux measurements
from different telescopes are consistent and there are obvious fluctuations of
the pulsar flux on a daily timescale. The largest variability was over a
factor of $\sim2$, between periastron and one week after, i.e.\ the third
and fourth \xmm/\nus\ observations. Over a longer timescale, the X-ray flux
gradually increased before periastron, then showed a drastic dip at periastron
and quick recovery, followed by a long declining trend by more than an order
of magnitude in a year \citep[see][]{ltn+18,cos+19}.

\subsection{Radio}
\begin{deluxetable}{lCCCC}
\tablewidth{0pt}
%\tabletypesize{\small}
\tablecaption{Radio Flux Densities of \psr\ Measured with the VLA
\label{tab:flux}}
\tablehead{
\colhead{Obs.\ Date} & \multicolumn{4}{c}{Flux Density\tablenotemark{a} (mJy)}\\
\cline{2-5} &\colhead{L band}&
\colhead{S band}&\colhead{C band}&\colhead{X band}}
\startdata
2017 Aug 14 & \nodata & 0.080\pm0.012 & \nodata & \nodata \\
2017 Sep 03 & 0.18\pm0.05 & 0.09\pm0.02 & < 0.05 & < 0.05 \\
2017 Oct 24 & 0.17\pm0.05 & < 0.05 & < 0.04 & < 0.04 \\
2017 Nov 07 & 0.17\pm0.04 & 0.033\pm0.016 & 0.030\pm0.014 & 0.035\pm0.014 \\
2017 Nov 13 & 0.08\pm0.06 & 0.040\pm0.019 & 0.050\pm0.015 & < 0.05 \\
2017 Nov 21 & 1.17\pm0.05 & 0.776\pm0.016 & 0.490\pm0.013 & 0.345\pm0.011 \\
2017 Dec 03 & 0.15\pm0.06 & 0.146\pm0.018 & 0.091\pm0.014 & 0.059\pm0.012 
\enddata 
\tablenotetext{a}{The uncertainties are estimated from the rms noise of the
intensity maps. In case of non-detection, 3$\sigma$ limits (i.e.\ three times
of the rms values), are reported. }
\end{deluxetable}

\begin{figure}[th]
\centering
\epsscale{0.5}
\plotone{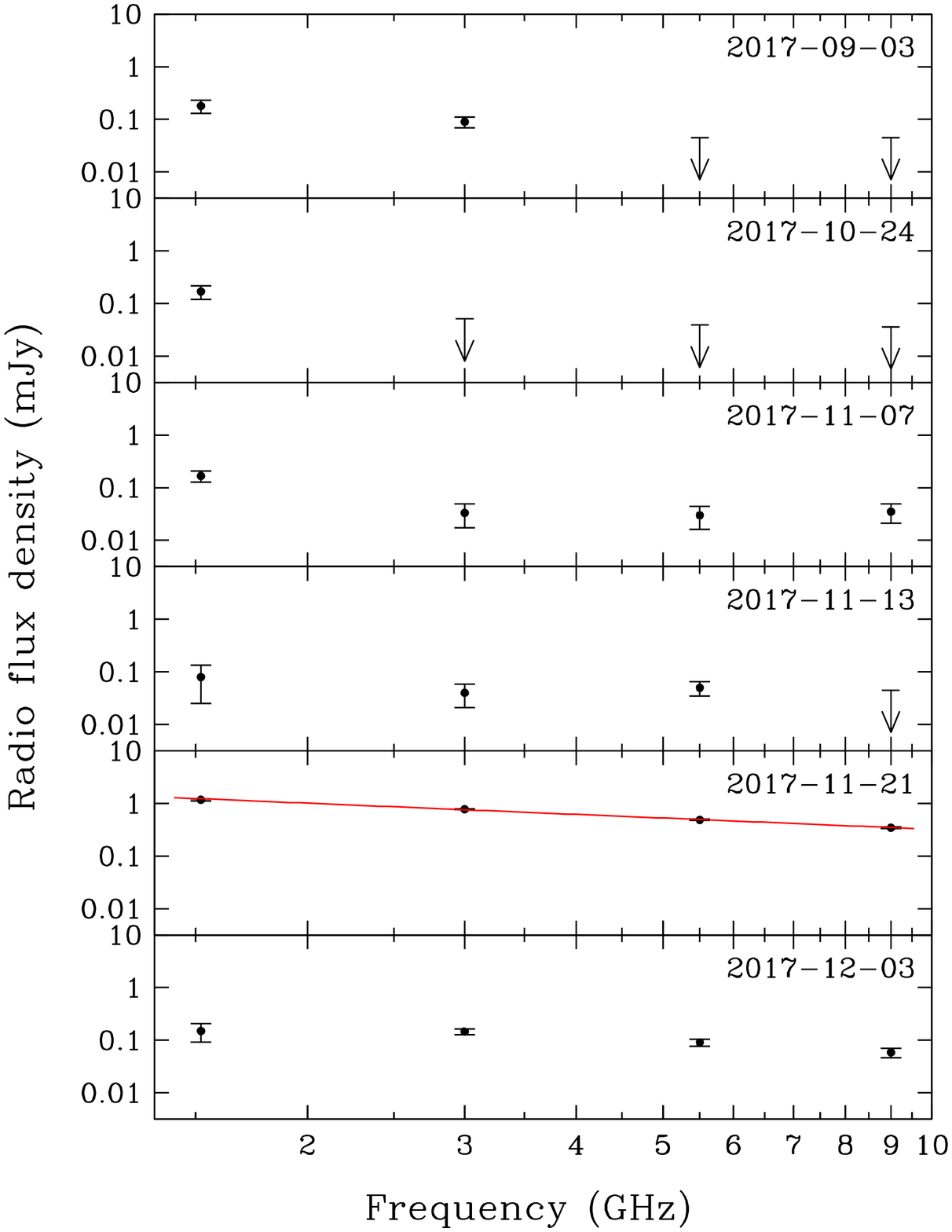}
\caption{Radio spectrum of \psr\ at different epochs measured with the VLA. In
cases of non-detection, 3$\sigma$ upper limits are plotted. The line shows the
best-fit spectrum of $S_\nu\propto\nu^{-0.71}$ when the emission was brightest.
\label{fig:rspec}}
\end{figure}

At radio frequencies, \psr\ was detected in all L-band VLA images, while the
emission sometimes fell below the observation sensitivity limits at higher
frequencies. The results are listed in Table~\ref{tab:flux} and plotted in
Figures~\ref{fig:comp} and \ref{fig:rspec}. No extended radio emission was
found, and \psr\ is consistent with a point source in all VLA images. Similar
to the X-ray emission, the radio flux exhibited a dip at periastron but
increased drastically by an order of magnitude one week after. In the last VLA
observation taken three weeks following periastron passage, the flux density
at the L band has returned to the pre-periastron level, but at higher
frequencies it remained about a factor of two higher. In the bottom panel of
Figure~\ref{fig:comp} we also plot the pulsed flux density of \psr\ at the L
band for comparison, which began to drop a few days after periastron and
stayed low for about 20\,days (Stappers et al.\ in prep.). Finally, we show in
Figure~\ref{fig:rspec} the radio spectrum at different epochs. When the
emission was brightest one week after periastron, its spectrum is well fit by
a power law with spectral index $\alpha=-0.71$ ($S_\nu\propto\nu^\alpha$). The
spectrum was however more complicated in other epochs and cannot be described
by a simple power law.

\section{Discussion}
Our new observations offered sensitive spectral measurements of \psr. We show
that the X-ray spectrum is well fit with a PL. The small photon index
($\Gamma<2$) rules out blackbody emission, indicating non-thermal distribution
of the emitting particles. Together with the lack of X-ray pulsations near
periastron and the strong flux variability, these suggest that the emission is
originated near the intra-binary shock between the stellar and pulsar winds,
and is likely to be synchrotron radiation from particles accelerated in the
shock. We showed that the X-ray PL spectrum flattened from $\Gamma\approx2$ to
1.2 at periastron. For synchrotron emission, this corresponded to a change of
particle distribution ${\rm d}N/{\rm d}E\propto E^{-p}$ from $p=3$ to 1.4. A
steeper spectrum indicates that the injected particles had sufficient time to
cool before escaping the radiation zone when the pulsar was far from
periastron. 

In the following discussion, we assume that MT91 213 has mass $M_\ast=15
\,M_\sun$ and radius $R_\ast=8\,R_\sun$ \citep{vgs96,hns10,hnl+17}.
Velocity of the radial wind is estimated to be
\begin{equation}
v_{\rm w}\sim\left(\frac{2GM_\ast}{R_\ast}\right)^{1/2}\approx850\,
\mathrm{km\,s^{-1}}~.
\end{equation}
We will take $v_{\rm w}=10^3$\kms\ for simplicity. Observations and
simulations suggest a mass loss rate of $\dot M_{\rm ej}\sim10^{-9}$--$10^{-8}
\,M_\sun$yr$^{-1}$ for Be stars \citep{sno81,pvn+08,krt14}, implying
wind density
\begin{equation}
\rho_0=\frac{\dot M_{\rm ej}}{4\pi R^2_\ast v_{\rm w}}
\sim10^{-16}-10^{-15}\,\mathrm{g\,cm^{-3}}~.
\end{equation}
If the stellar wind density decreases with distance $r$ from the star as
$\rho_{\rm w}(r)=\rho_0(R_\ast/r)^2$ \citep[e.g.,][]{pvc+18}, 
we can determine the wind shock distance $r_{\rm sh}$ from the pulsar by
pressure balance 
\begin{equation} 
\frac{\dot E}{4\pi c r_{\rm sh}^2} = \rho_w v_{\rm w}^2
= \rho_0\left(\frac{R_\ast}{r_p-r_{\rm sh}}\right)^2 v_{\rm w}^2~,
\end{equation}
where $\dot E=1.6\times 10^{35}$\ergs\ is the pulsar spin-down luminosity and
$r_p$ is the periastron separation. Using the typical neutron star mass
$M_{\rm NS}= 1.35\,M_\odot$ and the updated timing solution with orbital
period of 19113.4\,d and eccentricity of 0.9799 (Stappers et al.\ in prep.), we
obtain $r_p=0.71$\,AU. Solving the equation above gives
\begin{equation}
r_{\rm sh} = 2.9\times10^7\,\mathrm{km}\left(\frac{10^{-15}\,\mathrm{g\,cm^{-3}}}
{\rho_0}\right)^{1/2}\left(\frac{10^3\,\mathrm{km\,s^{-1}}}{v_{\rm w}}\right)~.
\end{equation}
This is much larger than the Bondi radius for accretion
\begin{equation}
r_{\rm B} \approx \frac{2GM_{\rm NS}}{v_{\rm w}^2} = 3.7\times10^5\,\mathrm{km}
\left(\frac{10^3\,\mathrm{km\,s^{-1}}}{v_{\rm w}}\right)^2~.
\end{equation}
Here we took the wind velocity as the relative velocity, since the neutron
star orbital velocity is always smaller than $v_{\rm w}$ even at periastron.
We therefore do not expect significant accretion.

The \xmm+\nus\ spectral fits indicate a possible break of the PL spectrum at
$\sim$5\,keV. This can be attributed to synchrotron cooling. Assuming energy
equipartition, the magnetic field of the wind at the shock is
\begin{equation}
B \sim (4\pi\rho_{\rm w}v_{\rm w}^2)^{1/2}
%   &= &11\,\mathrm{G}\left(
%\frac{\rho_{\rm w}}{10^{-15}\,\mathrm{g\,cm^{-3}}}\right)^{1/2}
 %\left(\frac{v_{\rm w}}{10^3\,\mathrm{km\,s^{-1}}}\right)\nonumber\\
= 0.8\,\mathrm{G}\left(\frac{\rho_0}{10^{-15}\,\mathrm{g\,cm^{-3}}}
 \right)^{1/2}\left(\frac{v_{\rm w}}{10^3\,\mathrm{km\,s^{-1}}}\right)~.
\end{equation}
The post-shock particles are expected to have a flow speed of $c/3$
\citep{rg74}. If we take their travel time across the periastron distance as
the synchrotron cooling time, we obtain a spectral break at $\sim$10\,keV,
comparable to our result. Finally, our new X-ray observations found an
absorption column density $N_{\rm H}=1.1\times10^{22}$\cms. This is higher
than the value $7.7\times 10^{21}$\cms\ inferred from the color excess of MT91
213 \citep{crr+09} and $3.4\times 10^{21}$\cms\ from the pulsar dispersion
measure \citep{hnk13}. We speculate that part of the absorption could be
contributed by the stellar wind
\begin{equation}
\Delta N_{\rm H}\sim\beta r_{\rm sh}\rho_{\rm w}/m_{\rm p}~,
\end{equation}
where $\beta$ parameterizes the density enhancement due to the shock and
$m_{\rm p}$ is hydrogen mass. For $\beta=10^3$, $\Delta N_{\rm H}=8\times
10^{21}$\cms, which could help explain the observed value.

We can compare \psr\ with a similar system \object{PSR B1259$-$63}/LS 2883
(hereafter B1259), the only other known gamma-ray binary consisting of a
pulsar and a Be-type star \citep{jml+92}. It has a comparable periastron
separation of 0.77\,AU but a much shorter orbital period of 3.4\,yr
\citep{nrh+11,mds+18}. The X-ray light curve of B1259 peaked $\sim$20 days
before and after periastron, corresponding to the times when the pulsar
crossed the Be stellar disk \citep{can+14,cnv+15}. This is different from the
slowly rising peak of \psr\ before periastron and the short-timescale flare
after. It was argued that only the latter was caused by the pulsar passing
through the disk \citep{ltn+18}. On the other hand, both systems exhibited
similar spectral variations. B1259 was also found to have \nh\ enhancement and
hardened, with $\Gamma$ changing from $\sim2$ to 1.5, near periastron
\citep{cnl+06,cnv+15}. High resolution \cxo\ observations revealed an extended
X-ray feature moving away from B1259 at $2\farcs0\pm0\farcs3\,$yr$^{-1}$,
corresponding to projected speed of 0.07$c$ \citep{kpd+14,phk+15}, and it can
be attributed to the pulsar and stellar wind mixture ejected near periastron
\citep{bb16}. For \psr, we did not find such extended emission in the entire
2\arcmin\ \cxo\ field of view, although our source is at a closer
distance \citep[1.39\,kpc versus 2.6\,kpc;][]{mds+18}.

\psr\ was previously claimed to show a hint of spectral variation, with
$\Delta N_{\rm H}\approx2 \times10^{22}$\cms\ and $\Delta \Gamma\approx0.8$
over two months in late 2016, by comparing between two different telescope
observations \citep{lkt+17}. In our study, we performed multiple observations
using a single set of instrument but did not find such large variations in our
observing period. As Figure~\ref{fig:xspec_t} indicates, the \xmm\ and \nus\
flux varied by a factor of two between the high and low states near
periastron, but the spectral parameters remained almost unchanged. This
suggests that the previous finding could be due to cross-calibration issues.

In the radio band, \citet{ltn+18} reported the detection of \psr\ with the VLA
in Aug 2017, and the emission was suggested to originate from the pulsar
magnetosphere. Our new observations support this picture. The VLA flux at the
L band stayed roughly constant before periastron and was consistent with the
pulsed value. At higher frequency, we argue that the pulsar wind emission
could be relatively important, due to its flat spectrum \citep[e.g.,]{krh06}.
Its flux is also expected to vary along the binary orbit. When combining with
the magnetospheric emission, these could possibly result in complicated
spectrum as shown in Figure~\ref{fig:rspec}.

\begin{figure}[th]
\centering
\epsscale{0.5}
\plotone{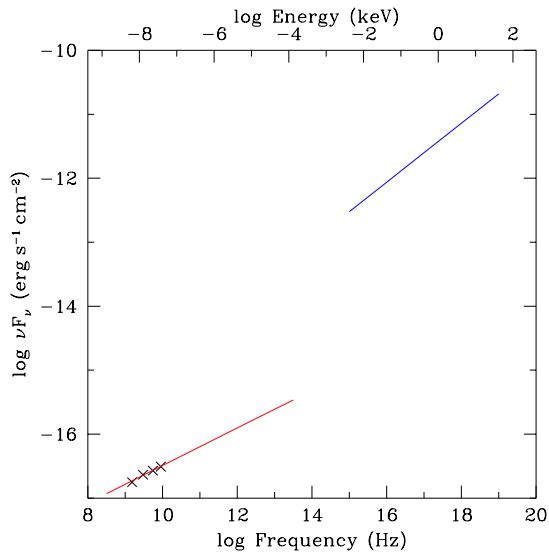}
\caption{Spectral-energy distribution of \psr\ during the X-ray and radio
flare that occurred one week after periastron. The best-fit simple PL spectra to
the radio and X-ray data are shown by the red and blue lines, respectively.
The data points indicate the VLA measurements.\label{fig:sed}}
\end{figure}

A strong unpulsed emission component emerged about one week after periastron.
It has flux density ten times higher than the pulsed component and fast rising
and decay times of less than a week each. We therefore consider it as a radio
flare. Figure~\ref{fig:comp} shows that the flare coincided with the X-ray
flare but decayed faster and nearly returned to the pre-periastron level after
a week. The radio flare has a spectral index of $\alpha=-0.71$, suggesting
that it could be synchrotron radiation \citep[see][]{jbw+05}. However, this is
steeper than the X-ray spectrum obtained with a single PL ($\alpha_X=1-\Gamma
=-0.54$) fit and comparable to the high energy part of the broken PL fit
($1-\Gamma_2=-0.65$). We plot in Figure~\ref{fig:sed} the broadband
spectral-energy distribution of \psr\ during the flare. It is clear that the
radio and X-ray spectra are not connected, suggesting that the two emission
could arise from different particle populations.

The radio behavior of \psr\ is very similar to that of B1259. Radio pulsations
from the latter also ceased near periastron. This was suggested to be due to
free-free absorption when the pulsar was behind the equatorial disk
\citep{mjm95}. The radio and X-ray light curves of B1259 showed a good
correlation \citep{can+14}, with an unpulsed radio emission component which
emerged during the X-ray flares before and after periastron. This emission has
a radio spectral index of $\alpha=-0.6$, comparable to $-0.71$ for \psr\
\citep{jbw+05}. It was believed to originate from electrons accelerated by the
intra-binary shock when B1259 moved through the stellar disk \citep{bmj+99}.
For the case of \psr, although the disk of MT91~213 shrank near periastron
such that it may not have directly interacted with the pulsar
\citep[see][]{csh+17,kel+17}, we argue that the pulsar wind shock could
possibly extend far enough to encounter the disk, or that the particle
environment near the disk was sufficiently different.

As a final remark, \citet{ltn+18} present the GeV light curve of \psr\ and
show that there is not a noticeable flare coinciding with the X-ray/radio
flare we found at about one week post-periastron.  However, it is worth noting
that they used two-week bins for the light curve and thus are not able to
resolve a GeV flare if one took place. On the other hand, such a GeV flare in
J2032 would behave very differently from the very luminous and variable GeV
flares of B1259, which take place around 20 days pre and post-periastron
\citep[see e.g.,][]{jwk+18,thp+18}.

\section{Conclusion}
We studied the X-ray and radio emission properties of \psr\ near the
periastron passage in late 2017. Using new observations taken with \cxo, \xmm,
and \nus, we showed that the X-ray emission of the system is non-thermal and
it can be described by a PL with a possible break around a few keV. We found
clear evidence of spectral variation in X-rays. The PL became harder at
periastron, and there is a hint of increasing column density. We found no
X-ray pulsations or short timescale flux variabilities. In the radio band, we
carried out VLA observations from 1 to 10\,GHz and discovered an unpulsed
emission component emerged one week after periastron, at the same time as the
X-ray flare. The spectral index of the emission suggests that it is
synchrotron emission, likely resulting from the interaction of the pulsar with
the Be stellar disk. For further investigation, detailed modeling is needed to
quantitatively explain the observed flux and spectral behavior. This will
provide insights into the particle acceleration process of the intra-binary
shock.

\acknowledgements
We thanks the referee for careful reading and useful suggestions.
C.Y.N. is supported by a GRF grant from the Hong Kong Government under HKU
17305416P. J.P.H. and E.V.G. acknowledge support from NASA through XMM-Newton
grant 80NSSC18K0402.  This work was also supported by \cxo\ Awards SAO
GO7-18055X (J.P.H.) and GO8-19123X (W.C.G.H). \cxo\ grants are issued by the
\cxo\ X-ray Center (CXC), which is operated by the Smithsonian Astrophysical
Observatory for and on behalf of NASA under contract NAS8-03060. Pulsar
observations with the Lovell Telescope are supported by a consolidated grant
from the STFC. Work at NRL is supported by NASA.

This research is in part based on the data obtained from the \cxo\ Data
Archive, and has made use of software provided by the CXC in the application
packages CIAO, ChIPS, and Sherpa. This research has used the observations
obtained with \xmm\ and ESA science mission with instruments and contributions
directly funded by the ESA member states and NASA. This research has also made
use of data obtained with \nus, a project led by Caltech, funded by NASA and
managed by NASA/JPL, and has utilized the NUSTARDAS software package, jointly
developed by the ASDC (Italy) and Caltech (USA). The National Radio Astronomy
Observatory is a facility of the National Science Foundation operated under
cooperative agreement by Associated Universities, Inc.

\facilities{CXO, XMM, NuSTAR, VLA}
\software{Sherpa \citep{fds01,dns+07}, XSPEC \citep{arn96}, CIAO
\citep[v4.10; ][]{fma+06}, ChIPS \citep{gmm+06}, NUSTARDAS, SAS
\cite[v15; ][]{gdf+04}, CASA \cite[v5.1; ][]{mws+07}} 

\appendix
\section{Net Count Rates of Nearby X-ray Sources\label{sec:bg}} 
\begin{figure}[ht]
\centering
\epsscale{0.4}
\plotone{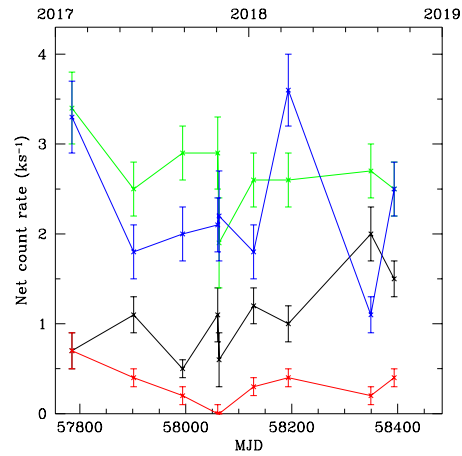}
\caption{Background-subtracted \cxo\ count rates of sources near \psr\ in
0.5--8\,keV range. The red, green, blue, and black lines represent
MT91 216, MT91 217, MT91 221, and CXOU J203213.5+412711, respectively.
\label{fig:bgrate}}
\end{figure}

As shown in
Figure~\ref{fig:cxo}, there are four background X-ray sources within 5\arcsec\
of \psr, namely, \object{MT91 216}, \object{MT91 217} (also known as
\object{Cyg OB2 4} and \object{BD+40\arcdeg\ 4219}), \object{MT91 221}, and
\object{CXOU J203213.5+412711}. These sources are covered by all
\cxo\ observations. We estimated their counts from 2\arcsec\ radius apertures
and then subtracted background from nearby source free regions. The resulting
net count rates are listed in Table~\ref{tab:bg} and plotted in
Figure~\ref{fig:bgrate}. These sources showed significant X-ray variabilities,
but they were all much fainter than \psr\ near periastron passage. Therefore
contamination in the \xmm\ and \nus\ spectra should be minimal. We did not
perform detailed spectral analysis as there are too few counts.

\begin{figure}[ht]
\centering
\epsscale{0.4}
\plotone{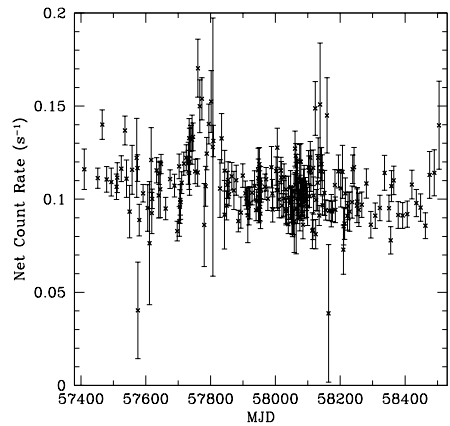}
\caption{Background-subtracted \sw\ count rates of V729 Cyg
in the 0.2--10\,keV energy range.\label{fig:bd}}
\end{figure}

We also report the X-ray light curve of a colliding wind binary, \object{V729
Cyg} (\object{BD+40\arcdeg\ 4220}, \object{Cyg OB2 5}, \object{Schulte 5}),
which is covered by the \sw\ field of view. We extracted the count rate from
30\arcsec\ radius source and background regions. The net count rate in
0.2--10\,keV energy range is plotted in Figure~\ref{fig:bd}. While this
eclipsing binary has an orbital period of 6.6\,d \citep{hal74}, our X-ray
light curve shows no obvious modulation in the 2--100\,d range. We can place
an upper limit of 1.4\% for the amplitude of any variation based on a simple
sinusoidal fit to the folded light curve at this precise period.

\begin{deluxetable}{lLLLL}
\tablewidth{0pt}
\tablecaption{Background-Subtracted \cxo\ Count Rates of MT91 216, MT91 217,
MT91 221, and CXOU J203213.5+412711 in 0.5--8\,keV \label{tab:bg}}
\tablehead{\colhead{Obs.\ Date}&\multicolumn{4}{c}{Net Count Rate (ks$^{-1}$)}\\
\cline{2-5} & \colhead{J203213.5} & \colhead{MT91 216}&
\colhead{MT91 217}& \colhead{MT91 221}}
\startdata
2017 Feb 01 &0.7\pm0.2&0.7\pm0.2&3.4\pm0.4&3.3\pm0.4\\
2017 May 28 &1.1\pm0.2&0.4\pm0.1&2.5\pm0.3&1.8\pm0.3\\
2017 Aug 29 &0.5\pm0.1&0.2\pm0.1&2.9\pm0.3&2.0\pm0.3\\
2017 Nov 03 &1.1\pm0.3&0.0\pm0.1&2.9\pm0.4&2.1\pm0.3\\
2017 Nov 05 &0.6\pm0.3&0.0\pm0.0&1.9\pm0.5&2.2\pm0.5\\
2018 Jan 09 &1.2\pm0.2&0.3\pm0.1&2.6\pm0.3&1.8\pm0.3\\
2018 Mar 16 &1.0\pm0.2&0.4\pm0.1&2.6\pm0.3&3.6\pm0.4\\
2018 Aug 19 &2.0\pm0.3&0.2\pm0.1&2.7\pm0.3&1.1\pm0.2\\
2018 Oct 02 &1.5\pm0.2&0.4\pm0.1&2.5\pm0.3&2.5\pm0.3
\enddata 
\end{deluxetable}

\clearpage

\bibliography{ncy}

\begin{thebibliography}{}
\expandafter\ifx\csname natexlab\endcsname\relax\def\natexlab#1{#1}\fi
\providecommand{\url}[1]{\href{#1}{#1}}
\providecommand{\dodoi}[1]{doi:~\href{http://doi.org/#1}{\nolinkurl{#1}}}
\providecommand{\doeprint}[1]{\href{http://ascl.net/#1}{\nolinkurl{http://ascl.net/#1}}}
\providecommand{\doarXiv}[1]{\href{https://arxiv.org/abs/#1}{\nolinkurl{https://arxiv.org/abs/#1}}}

\bibitem[{{Abdo} {et~al.}(2009){Abdo}, {Ackermann}, {Ajello}, {Anderson},
  {Atwood}, {Axelsson}, {Baldini}, {Ballet}, {Barbiellini}, {Baring},
  {Bastieri}, \& et~al.}]{aaa+09a}
{Abdo}, A.~A., {Ackermann}, M., {Ajello}, M., {et~al.} 2009, Science, 325, 840,
  \dodoi{10.1126/science.1175558}

\bibitem[{{Abeysekara} {et~al.}(2018){Abeysekara}, {Benbow}, {Bird}, {Brill},
  {Brose}, {Buckley}, {Chromey}, {Daniel}, {Falcone}, {Finley}, \&
  et~al.}]{abb+18}
{Abeysekara}, A.~U., {Benbow}, W., {Bird}, R., {et~al.} 2018, \apjl, 867, L19,
  \dodoi{10.3847/2041-8213/aae70e}

\bibitem[{{Aliu} {et~al.}(2014){Aliu}, {Aune}, {Behera}, {Beilicke}, {Benbow},
  {Berger}, {Bird}, {Bouvier}, {Buckley}, {Bugaev}, {Cerruti}, {Chen},
  {Ciupik}, {Connolly}, {Cui}, {Dumm}, {Dwarkadas}, {Errando}, {Falcone},
  {Federici}, {Feng}, {Finley}, {Fleischhack}, {Fortin}, {Fortson}, {Furniss},
  {Galante}, {Gillanders}, {Gotthelf}, {Griffin}, {Griffiths}, {Grube}, {Gyuk},
  {Hanna}, {Holder}, {Hughes}, {Humensky}, {Johnson}, {Kaaret}, {Kargaltsev},
  {Kertzman}, {Khassen}, {Kieda}, {Krennrich}, {Lang}, {Madhavan}, {Maier},
  {McArthur}, {McCann}, {Millis}, {Moriarty}, {Mukherjee}, {Nieto},
  {O'Faol{\'a}in de Bhr{\'o}ithe}, {Ong}, {Otte}, {Pandel}, {Park}, {Pohl},
  {Popkow}, {Prokoph}, {Quinn}, {Ragan}, {Rajotte}, {Reyes}, {Reynolds},
  {Richards}, {Roache}, {Roberts}, {Sembroski}, {Shahinyan}, {Smith},
  {Staszak}, {Telezhinsky}, {Tucci}, {Tyler}, {Vincent}, {Wakely}, {Weinstein},
  {Welsing}, {Wilhelm}, {Williams}, \& {Zitzer}}]{aab+14}
{Aliu}, E., {Aune}, T., {Behera}, B., {et~al.} 2014, \apj, 788, 78,
  \dodoi{10.1088/0004-637X/788/1/78}

\bibitem[{{Arnaud}(1996)}]{arn96}
{Arnaud}, K.~A. 1996, in Astronomical Society of the Pacific Conference Series,
  Vol. 101, Astronomical Data Analysis Software and Systems V, ed. G.~H.
  {Jacoby} \& J.~{Barnes}, 17

\bibitem[{{Ball} {et~al.}(1999){Ball}, {Melatos}, {Johnston}, \& {Skj{\ae}
  Raasen}}]{bmj+99}
{Ball}, L., {Melatos}, A., {Johnston}, S., \& {Skj{\ae} Raasen}, O. 1999,
  \apjl, 514, L39, \dodoi{10.1086/311928}

\bibitem[{{Barkov} \& {Bosch-Ramon}(2016)}]{bb16}
{Barkov}, M.~V., \& {Bosch-Ramon}, V. 2016, \mnras, 456, L64,
  \dodoi{10.1093/mnrasl/slv171}

\bibitem[{{Bednarek} {et~al.}(2018){Bednarek}, {Banasi{\'n}ski}, \&
  {Sitarek}}]{bbs18}
{Bednarek}, W., {Banasi{\'n}ski}, P., \& {Sitarek}, J. 2018, Journal of Physics
  G Nuclear Physics, 45, 015201, \dodoi{10.1088/1361-6471/aa97ee}

\bibitem[{{Camilo} {et~al.}(2009){Camilo}, {Ray}, {Ransom}, {Burgay},
  {Johnson}, {Kerr}, {Gotthelf}, {Halpern}, {Reynolds}, {Romani}, {Demorest},
  {Johnston}, {van Straten}, {Saz Parkinson}, {Ziegler}, {Dormody}, {Thompson},
  {Smith}, {Harding}, {Abdo}, {Crawford}, {Freire}, {Keith}, {Kramer},
  {Roberts}, {Weltevrede}, \& {Wood}}]{crr+09}
{Camilo}, F., {Ray}, P.~S., {Ransom}, S.~M., {et~al.} 2009, \apj, 705, 1,
  \dodoi{10.1088/0004-637X/705/1/1}

\bibitem[{{Chernyakova} {et~al.}(2006){Chernyakova}, {Neronov}, {Lutovinov},
  {Rodriguez}, \& {Johnston}}]{cnl+06}
{Chernyakova}, M., {Neronov}, A., {Lutovinov}, A., {Rodriguez}, J., \&
  {Johnston}, S. 2006, \mnras, 367, 1201,
  \dodoi{10.1111/j.1365-2966.2005.10039.x}

\bibitem[{{Chernyakova} {et~al.}(2014){Chernyakova}, {Abdo}, {Neronov},
  {McSwain}, {Mold{\'o}n}, {Rib{\'o}}, {Paredes}, {Sushch}, {de Naurois},
  {Schwanke}, {Uchiyama}, {Wood}, {Johnston}, {Chaty}, {Coleiro}, {Malyshev},
  \& {Babyk}}]{can+14}
{Chernyakova}, M., {Abdo}, A.~A., {Neronov}, A., {et~al.} 2014, \mnras, 439,
  432, \dodoi{10.1093/mnras/stu021}

\bibitem[{{Chernyakova} {et~al.}(2015){Chernyakova}, {Neronov}, {van Soelen},
  {Callanan}, {O'Shaughnessy}, {Babyk}, {Tsygankov}, {Vovk}, {Krivonos},
  {Tomsick}, {Malyshev}, {Li}, {Wood}, {Torres}, {Zhang}, {Kretschmar},
  {McSwain}, {Buckley}, \& {Koen}}]{cnv+15}
{Chernyakova}, M., {Neronov}, A., {van Soelen}, B., {et~al.} 2015, \mnras, 454,
  1358, \dodoi{10.1093/mnras/stv1988}

\bibitem[{{Coe} {et~al.}(2017){Coe}, {Steele}, {Ho}, {Stappers}, {Lyne},
  {Halpern}, {Ray}, {Johnson}, {Ng}, \& {Kerr}}]{csh+17}
{Coe}, M.~J., {Steele}, I.~A., {Ho}, W.~C.~G., {et~al.} 2017, The Astronomer's
  Telegram, 10920

\bibitem[{{Coe} {et~al.}(2019){Coe}, {Okazaki}, {Steele}, {Ng}, {Ho}, {Lyne},
  {Stappers}, {Johnson}, {Ray}, \& {Kerr}}]{cos+19}
{Coe}, M.~J., {Okazaki}, A.~T., {Steele}, I.~A., {et~al.} 2019, \mnras, 485,
  1864, \dodoi{10.1093/mnras/stz515}

\bibitem[{{Doe} {et~al.}(2007){Doe}, {Nguyen}, {Stawarz}, {Refsdal},
  {Siemiginowska}, {Burke}, {Evans}, {Evans}, {McDowell}, {Houck}, \&
  {Nowak}}]{dns+07}
{Doe}, S., {Nguyen}, D., {Stawarz}, C., {et~al.} 2007, in Astronomical Society
  of the Pacific Conference Series, Vol. 376, Astronomical Data Analysis
  Software and Systems XVI, ed. R.~A. {Shaw}, F.~{Hill}, \& D.~J. {Bell}, 543

\bibitem[{{Dubus}(2013)}]{dub13}
{Dubus}, G. 2013, \aapr, 21, 64, \dodoi{10.1007/s00159-013-0064-5}

\bibitem[{{Dubus} {et~al.}(2017){Dubus}, {Guillard}, {Petrucci}, \&
  {Martin}}]{dgp+17}
{Dubus}, G., {Guillard}, N., {Petrucci}, P.-O., \& {Martin}, P. 2017, \aap,
  608, A59, \dodoi{10.1051/0004-6361/201731084}

\bibitem[{{Freeman} {et~al.}(2001){Freeman}, {Doe}, \& {Siemiginowska}}]{fds01}
{Freeman}, P., {Doe}, S., \& {Siemiginowska}, A. 2001, in SPIE, Vol. 4477,
  Astronomical Data Analysis, ed. J.-L. {Starck} \& F.~D. {Murtagh}, 76

\bibitem[{{Fruscione} {et~al.}(2006){Fruscione}, {McDowell}, {Allen},
  {Brickhouse}, {Burke}, {Davis}, {Durham}, {Elvis}, {Galle}, {Harris},
  {Huenemoerder}, {Houck}, {Ishibashi}, {Karovska}, {Nicastro}, {Noble},
  {Nowak}, {Primini}, {Siemiginowska}, {Smith}, \& {Wise}}]{fma+06}
{Fruscione}, A., {McDowell}, J.~C., {Allen}, G.~E., {et~al.} 2006, in SPIE,
  Vol. 6270, Observatory Operations: Strategies, Processes, and Systems, ed.
  D.~R. {Silva} \& D.~R. E., 62701V

\bibitem[{{Gabriel} {et~al.}(2004){Gabriel}, {Denby}, {Fyfe}, {Hoar}, {Ibarra},
  {Ojero}, {Osborne}, {Saxton}, {Lammers}, \& {Vacanti}}]{gdf+04}
{Gabriel}, C., {Denby}, M., {Fyfe}, D.~J., {et~al.} 2004, in \aspconf, Vol.
  314, Astronomical Data Analysis Software and Systems (ADASS) XIII, ed.
  F.~{Ochsenbein}, M.~G. {Allen}, \& D.~{Egret}, 759

\bibitem[{{Germain} {et~al.}(2006){Germain}, {Milaszewski}, {McLaughlin},
  {Miller}, {Evans}, {Evans}, \& {Burke}}]{gmm+06}
{Germain}, G., {Milaszewski}, R., {McLaughlin}, W., {et~al.} 2006, in
  Astronomical Society of the Pacific Conference Series, Vol. 351, Astronomical
  Data Analysis Software and Systems XV, ed. C.~{Gabriel}, C.~{Arviset},
  D.~{Ponz}, \& S.~{Enrique}, 57

\bibitem[{{Hall}(1974)}]{hal74}
{Hall}, D.~S. 1974, \actaa, 24, 69

\bibitem[{{He} {et~al.}(2013){He}, {Ng}, \& {Kaspi}}]{hnk13}
{He}, C., {Ng}, C.-Y., \& {Kaspi}, V.~M. 2013, \apj, 768, 64,
  \dodoi{10.1088/0004-637X/768/1/64}

\bibitem[{{Ho} {et~al.}(2017){Ho}, {Ng}, {Lyne}, {Stappers}, {Coe}, {Halpern},
  {Johnson}, \& {Steele}}]{hnl+17}
{Ho}, W.~C.~G., {Ng}, C.-Y., {Lyne}, A.~G., {et~al.} 2017, \mnras, 464, 1211,
  \dodoi{10.1093/mnras/stw2420}

\bibitem[{{Hohle} {et~al.}(2010){Hohle}, {Neuh{\"a}user}, \& {Schutz}}]{hns10}
{Hohle}, M.~M., {Neuh{\"a}user}, R., \& {Schutz}, B.~F. 2010, Astronomische
  Nachrichten, 331, 349, \dodoi{10.1002/asna.200911355}

\bibitem[{{Jennings} {et~al.}(2018){Jennings}, {Kaplan}, {Chatterjee},
  {Cordes}, \& {Deller}}]{jkc+18}
{Jennings}, R.~J., {Kaplan}, D.~L., {Chatterjee}, S., {Cordes}, J.~M., \&
  {Deller}, A.~T. 2018, \apj, 864, 26, \dodoi{10.3847/1538-4357/aad084}

\bibitem[{{Johnson} {et~al.}(2018){Johnson}, {Wood}, {Kerr}, {Corbet},
  {Cheung}, {Ray}, \& {Omodei}}]{jwk+18}
{Johnson}, T.~J., {Wood}, K.~S., {Kerr}, M., {et~al.} 2018, \apj, 863, 27,
  \dodoi{10.3847/1538-4357/aad185}

\bibitem[{{Johnston} {et~al.}(2005){Johnston}, {Ball}, {Wang}, \&
  {Manchester}}]{jbw+05}
{Johnston}, S., {Ball}, L., {Wang}, N., \& {Manchester}, R.~N. 2005, \mnras,
  358, 1069, \dodoi{10.1111/j.1365-2966.2005.08854.x}

\bibitem[{{Johnston} {et~al.}(1992){Johnston}, {Manchester}, {Lyne}, {Bailes},
  {Kaspi}, {Qiao}, \& {D'Amico}}]{jml+92}
{Johnston}, S., {Manchester}, R.~N., {Lyne}, A.~G., {et~al.} 1992, \apjl, 387,
  L37, \dodoi{10.1086/186300}

\bibitem[{{Kargaltsev} {et~al.}(2014){Kargaltsev}, {Pavlov}, {Durant},
  {Volkov}, \& {Hare}}]{kpd+14}
{Kargaltsev}, O., {Pavlov}, G.~G., {Durant}, M., {Volkov}, I., \& {Hare}, J.
  2014, \apj, 784, 124, \dodoi{10.1088/0004-637X/784/2/124}

\bibitem[{{Kaspi} {et~al.}(2006){Kaspi}, {Roberts}, \& {Harding}}]{krh06}
{Kaspi}, V.~M., {Roberts}, M.~S.~E., \& {Harding}, A.~K. 2006, in Compact
  Stellar X-ray Sources, ed. W.~H.~G. {Lewin} \& M.~{van der Klis} (Cambridge:
  Cambridge University Press), 279

\bibitem[{{Kolka} {et~al.}(2017){Kolka}, {Eenm{\"a}e}, {Laur}, \&
  {Aret}}]{kel+17}
{Kolka}, I., {Eenm{\"a}e}, T., {Laur}, J., \& {Aret}, A. 2017, Research Notes
  of the American Astronomical Society, 1, 37, \dodoi{10.3847/2515-5172/aa9f17}

\bibitem[{{Krti{\v c}ka}(2014)}]{krt14}
{Krti{\v c}ka}, J. 2014, \aap, 564, A70, \dodoi{10.1051/0004-6361/201321980}

\bibitem[{{Li} {et~al.}(2017){Li}, {Kong}, {Tam}, {Hou}, {Takata}, \&
  {Hui}}]{lkt+17}
{Li}, K.~L., {Kong}, A.~K.~H., {Tam}, P.~H.~T., {et~al.} 2017, \apj, 843, 85,
  \dodoi{10.3847/1538-4357/aa784e}

\bibitem[{{Li} {et~al.}(2018){Li}, {Takata}, {Ng}, {Kong}, {Tam}, {Hui}, \&
  {Cheng}}]{ltn+18}
{Li}, K.~L., {Takata}, J., {Ng}, C.~W., {et~al.} 2018, \apj, 857, 123,
  \dodoi{10.3847/1538-4357/aab848}

\bibitem[{{Lyne} {et~al.}(2015){Lyne}, {Stappers}, {Keith}, {Ray}, {Kerr},
  {Camilo}, \& {Johnson}}]{lsk+15}
{Lyne}, A.~G., {Stappers}, B.~W., {Keith}, M.~J., {et~al.} 2015, \mnras, 451,
  581, \dodoi{10.1093/mnras/stv236}

\bibitem[{{McMullin} {et~al.}(2007){McMullin}, {Waters}, {Schiebel}, {Young},
  \& {Golap}}]{mws+07}
{McMullin}, J.~P., {Waters}, B., {Schiebel}, D., {Young}, W., \& {Golap}, K.
  2007, in \aspconf, Vol. 376, Astronomical Data Analysis Software and Systems
  XVI, ed. R.~A. {Shaw}, F.~{Hill}, \& D.~J. {Bell}, 127

\bibitem[{{Melatos} {et~al.}(1995){Melatos}, {Johnston}, \& {Melrose}}]{mjm95}
{Melatos}, A., {Johnston}, S., \& {Melrose}, D.~B. 1995, \mnras, 275, 381,
  \dodoi{10.1093/mnras/275.2.381}

\bibitem[{{Miller-Jones} {et~al.}(2018){Miller-Jones}, {Deller}, {Shannon},
  {Dodson}, {Mold{\'o}n}, {Rib{\'o}}, {Dubus}, {Johnston}, {Paredes}, {Ransom},
  \& {Tomsick}}]{mds+18}
{Miller-Jones}, J.~C.~A., {Deller}, A.~T., {Shannon}, R.~M., {et~al.} 2018,
  \mnras, 479, 4849, \dodoi{10.1093/mnras/sty1775}

\bibitem[{{Mukherjee} {et~al.}(2007){Mukherjee}, {Gotthelf}, \&
  {Halpern}}]{mgh07}
{Mukherjee}, R., {Gotthelf}, E.~V., \& {Halpern}, J.~P. 2007, \apss, 309, 29,
  \dodoi{10.1007/s10509-007-9452-5}

\bibitem[{{Negueruela} {et~al.}(2011){Negueruela}, {Rib{\'o}}, {Herrero},
  {Lorenzo}, {Khangulyan}, \& {Aharonian}}]{nrh+11}
{Negueruela}, I., {Rib{\'o}}, M., {Herrero}, A., {et~al.} 2011, \apjl, 732,
  L11, \dodoi{10.1088/2041-8205/732/1/L11}

\bibitem[{{Pavlov} {et~al.}(2015){Pavlov}, {Hare}, {Kargaltsev}, {Rangelov}, \&
  {Durant}}]{phk+15}
{Pavlov}, G.~G., {Hare}, J., {Kargaltsev}, O., {Rangelov}, B., \& {Durant}, M.
  2015, \apj, 806, 192, \dodoi{10.1088/0004-637X/806/2/192}

\bibitem[{{Petropoulou} {et~al.}(2018){Petropoulou}, {Vasilopoulos},
  {Christie}, {Giannios}, \& {Coe}}]{pvc+18}
{Petropoulou}, M., {Vasilopoulos}, G., {Christie}, I.~M., {Giannios}, D., \&
  {Coe}, M.~J. 2018, \mnras, 474, L22, \dodoi{10.1093/mnrasl/slx185}

\bibitem[{{Puls} {et~al.}(2008){Puls}, {Vink}, \& {Najarro}}]{pvn+08}
{Puls}, J., {Vink}, J.~S., \& {Najarro}, F. 2008, \aapr, 16, 209,
  \dodoi{10.1007/s00159-008-0015-8}

\bibitem[{{Rees} \& {Gunn}(1974)}]{rg74}
{Rees}, M.~J., \& {Gunn}, J.~E. 1974, \mnras, 167, 1

\bibitem[{{Snow}(1981)}]{sno81}
{Snow}, Jr., T.~P. 1981, \apj, 251, 139, \dodoi{10.1086/159448}

\bibitem[{{Takata} {et~al.}(2017){Takata}, {Tam}, {Ng}, {Li}, {Kong}, {Hui}, \&
  {Cheng}}]{ttn+17}
{Takata}, J., {Tam}, P.~H.~T., {Ng}, C.~W., {et~al.} 2017, \apj, 836, 241,
  \dodoi{10.3847/1538-4357/aa5c80}

\bibitem[{{Tam} {et~al.}(2018){Tam}, {He}, {Pal}, \& {Cui}}]{thp+18}
{Tam}, P.~H.~T., {He}, X.~B., {Pal}, P.~S., \& {Cui}, Y. 2018, \apj, 862, 165,
  \dodoi{10.3847/1538-4357/aacf00}

\bibitem[{{Vacca} {et~al.}(1996){Vacca}, {Garmany}, \& {Shull}}]{vgs96}
{Vacca}, W.~D., {Garmany}, C.~D., \& {Shull}, J.~M. 1996, \apj, 460, 914,
  \dodoi{10.1086/177020}

\bibitem[{{Wilms} {et~al.}(2000){Wilms}, {Allen}, \& {McCray}}]{wam00}
{Wilms}, J., {Allen}, A., \& {McCray}, R. 2000, \apj, 542, 914,
  \dodoi{10.1086/317016}

\end{thebibliography}
\end{document}